# The organization of a three-manual keyboard for 53-tone tempered and other tempered systems.

**Burskii V.P.**

*Moscow Institute for Physics and Technology; Institute of applied mathematics and mechanics of Nat. Ac. Sci. Ukraine, Donetsk*

# Организация трех-мануальной клавиатуры для 53-ступенной и других темперированных систем.

**Бурский В.П.**

*МФТИ; Институт прикладной математики и механики НАН Украины, Донецк*

**Abstract:** The aim is to explore new opportunities of the pitch organization of the musical scale. Specifically, a numerical comparison of the different musical temperaments among themselves in the degree of approximation of the Pythagorean scale is provided, and thus it numerically substantiates the thesis that the 53-tone tempered system is the most advanced among possible others. We present numerical data on the approximation of overtones from first twenty by steps of the 53-tone temperament. Here were proposed some schemes of the three-manual keyboard for the implementation of 53-tone temperament, which are also implemented at the same time for 12 -, 17 -, 24 -, 29 - and 41-sounding system. If there are technical means then these schemes can be used to play music in any temperaments, based on said number of steps.

**Реферат:** Целью работы является исследование новых возможностей высотной организации музыкального звукоряда. А именно, предлагаются схемы организации трех-мануальной клавиатуры для реализации 53-ступенной темперации, в которых также одновременно реализуются 12-, 17-, 24-, 29-, 41-звучные системы, каждая в своей схеме, и которые при наличии технических средств могут быть использованы при исполнении музыкальных произведений в любой из темпераций, основанных на указанных числах ступеней. В качестве введения и для обоснования тезиса о необходимости использования в музыкальной теории и практике иных, нежели 12-звучная, темпераций проводится численное сравнение различных музыкальных темпераций между собой в степени приближения пифагорейского звукоряда. Тем самым, численно обосновывается известный тезис о том, что 53-ступенная темперированная система является в этом плане наиболее совершенной среди возможных других. Приводятся также числовые данные о приближении обертонов первой двадцатки ступенями 53-ступенной темперации. В предлагаемых схемах организации звукоряда реализуются известные равенства 5+7=12, 12+5=17, 12+12=24, 12+17=29, 12+29=41, 12+17+24=53, возникающие при рассмотрении темперированных систем. Рассмотрены также некоторые связанные теоретические вопросы, например, кварто-квинтовый круг, системы интонирования и другие.

*1. Введение.* Развитие музыки приводит к естественному появлению и апробации новых изобразительных средств, в частности, систем расположения звуков по высоте. С историей этого направления в теории музыки читатель может познакомиться по книге [10], по брошюре [13], устоявшиеся представления изложены, например, в учебнике [8], современный взгляд на историю, нынешнее состояние и практику микрохроматики содержится в обзоре [5], связи с физиологией восприятия см., например, в книге [4] и статье [7]. В настоящей работе изучаются новые возможности высотной организации звукоряда в музыке. Актуальность вопроса сейчас часто подчеркивается в публикациях (см., напр., [6], [5], [2]).

В настоящей работе мы исходим из базового положения о том, что для удобства исполнения система расположения звуков должна быть равномерной, т.е. темперированной, и, как это принято со времен Пифагора, построенной на основе квинты. Ниже рассматриваются различные темперированные системы, приводится численное сравнение точности отображения в них основных обертонов и на этой основе обсуждаются преимущества перед другими 53-ступенной темперированной системы, которой в основном и посвящены дальнейшие построения. Преимущества 53-ступенной темперации (называемой иногда темперацией Меркатора) многократно отмечались в литературе, см., напр., книги [6], [13] (в работе [13], §6 они прямо называются «преимуществами чистого строя»). Более того, были предложены и изготовлены инструменты, реализующие эту темперацию. Так, в той же работе [13] упоминаются инструменты с 53-ступенной темперацией: у R.H.M. Bosanquet (Лондон, 1875) с одним мануалом и 84 клавишами в октаве, а также инструмент доктора Sh. Tanaka с 26 звуками в октаве и транспонирующим приспособлением (c ↔ ges). Более современные предложения организации клавиатуры для реализации 53-ступенной темперации имеются в работе [12], где клавиши в виде сот располагаются в несколько рядов. Приведем еще цитату из работы [5]: «об акустическом превосходстве этой темперации писал еще Г.М. Римский-Корсаков. Однако большой популярности эта темперация не завоевала, вероятно, в силу своей непрактичности».

Настоящая работа как раз направлена на улучшение практичности 53-ступенной системы. Принципиально новой здесь является предлагаемая ниже идея расположения ступеней в виде трёх-мануальной клавиатуры, реализация которой позволяет, с одной стороны, например, в теоретических изысканиях, использовать полный набор звуков системы, и, с другой стороны, при исполнении большинства произведений пользоваться двумя (из имеющихся трех) мануалами или даже привычным для всех одним фортепианным мануалом. Кроме того, предлагаемая трёх-мануальная клавиатура позволяет одновременно реализовать различные системы интонирования, а также 12-, 17-, 24-, 29- и 41-звучные темперированные системы при наличии соответствующих технических средств. Отметим еще, что трех-мануальная клавиатура для исполнения произведений в 53-ступенной системе в литературе до нас не предлагалась.

*2. Возможности приближения чистой (натуральной) квинты в различных темперированных системах.* Выберем какой-нибудь основной тон (фундаментальной) частоты $\omega_1$ (герц) и будем от него строить темперированную музыкальную систему. Напомним (см., напр., [1], а также [9]), что интервальный коэффициент k какого-либо тона частоты $\omega_2$ по отношению к основному тону (и интервальный коэффициент соответствующего интервала) выражается формулой $k = \omega_2 / \omega_1$, интервальный коэффициент чистой октавы равен 2, а интервальный коэффициент чистой (натуральной) квинты равен 3/2. Напомним еще, что высота этого тона с интервальным коэффициентом k по отношению к основному (и величина интервала между ними) характеризуется величиной $h=\{\log_2 k\}$, которую будем называть высотой этого тона (здесь $\{d\}$— дробная часть числа d, например, $\{3,14\}=0,14$). Напомним также, что для числовой характеризации тонов и соответствующих интервалов октаву по логарифмической шкале принято делить на 1200 частей. Эти части называются центами (обозначение c от англ. cent), тогда высота нашего тона (и величина интервала) в центах равна 1200 h. Тогда равномерная темперация – это задание тонов, разбивающих октаву на равные интервалы, т.е. каждый из этих интервалов содержит одно и то же число центов.

Среди множества всех интервалов между различными тонами темперации должны быть такие, которые достаточно хорошо приближают консонирующие интервалы, такие как чистая квинта и большая терция, порожденные соответственно 3-м и 5-м обертонами основного тона. Обычно между этими двумя возможностями выбирают лучшее приближение для квинты, нежели для терции, что ниже будем предполагать и мы тоже. В темперированном строе, состоящем из q ступеней, мы должны получить каждую из этих ступеней, равномерно двигаясь по шкале частот от коэффициента $1=2^0$ до октавного коэффициента $2=2^1$ шагами по $2^{1/q}$ (всего q ступеней в октаве). При этом чистую квинту должен приближать интервал между, например, первой ступенью с коэффициентом $1=2^0$ и p+1-й ступенью с коэффициентом $\beta = 2^{p/q}$ . Мы получили, что высота темперированного квинтового тона равна p/q (рациональному числу). А высота чистого квинтового тона выражается числом $\alpha = \log_2 3/2 = 0,5849625007$, что составляет 701,95500084 c.

Для того, чтобы темперация имела бы практический смысл и музыкальное звучание, важно, чтобы её темперированная квинта была бы близка к чистой квинте, поэтому следует рассматривать те системы, т.е. те целые числа q, для которых найдется целое p такое, чтобы число p/q было бы достаточно близко к числу $\alpha$. При этом, чем меньше их разность $\Delta = \alpha - p/q$ , тем ближе этот темперированный звукоряд к звукоряду Пифагора (расширенному диатоническому), т.е. составленному из чистых квинт, а потому нетемперируемому принципиально, поскольку $\alpha$ – иррациональное число. Подчеркнем, что дробь p/q равна показателю степени в выражении для интервального коэффициента $\beta = 2^{p/q}$ темперированной квинты, т.е. для отношения частоты темперированного квинтового тона к частоте основного тона в нашем рассмотрении. Ниже, в таблице 1, приведены лучшие рациональные приближения числа $\alpha$ и их отклонения $\Delta = \alpha - p/q$ ( $\alpha$ = 0,5849625007, $\alpha \cdot 1200$ = 701,95500084 ), причем, может быть, стоит отметить, что дробь 7/12 = =0,5833333333 ( = 700 c) отвечает случаю баховской 12-звучной темперации. Вычисления, представленные в таблице 1, не являются новыми, различные результаты многократно появлялись в литературе, а сама таблица 1, как и предыдущие пояснения, приведена здесь для удобства читателя. Заметим, что число ступеней в квинте мы понимаем здесь как номер ступени квинты минус один, то есть, если основной тон имеет номер 1, то квинтовый тон – номер p+1.

*Таблица 1. Приближения чистой квинты в различных темперациях*

| Число звуков в темперации q | Число ступеней в квинте p | Высота темперированной квинты, т.е. отношение p/q | Отношение p/q в центах, т.е. число p/q · 1200 | Отклонение $\Delta = \alpha - p/q$ в центах, т.е. $\Delta \cdot 1200$ |
|---|---|---|---|---|
| 5 | 3 | 0,6000000000 | 720,00000000 | −18,04499916 |
| 7 | 4 | 0,5714285714 | 685,71428568 | +16,24071516 |
| 12 | 7 | 0,5833333333 | 700,00000000 | + 1,95500084 |
| 17 | 10 | 0,5882352941 | 705,88235292 | − 3,92735208 |
| 21 | 12 | 0,5714285714 | 685,71428568 | +16,24071516 |
| 24 | 14 | 0,5833333333 | 700,00000000 | + 1,95500084 |
| 29 | 17 | 0,5862068966 | 703,44827592 | + 1,49327508 |
| 31 | 18 | 0,5806451613 | 696,77419356 | + 5,18080728 |
| 41 | 24 | 0,5853658537 | 702,43902444 | − 0,48402360 |
| 53 | 31 | 0,5849056604 | 701,88679248 | + 0,06820836 |
| 65 | 38 | 0,5846153846 | 701,53846152 | + 0,41653932 |
| 359 | 210 | 0,5849582172 | 701,94986064 | + 0,00514020 |

Отметим, что в выборе указанных в таблице чисел p и q нет ничего мистического. Рациональные числа p/q =7/12, 10/17, 17/29, 24/41, 31/53 возникают в математической теории приближения действительных чисел рациональными числами (что отмечено, например, в книге [11]). Они представляют собой так называемые подходящие дроби для числа $\alpha$, которые явно считаются по разложению числа $\alpha$ в непрерывную цепную дробь, заинтересованный читатель сможет найти простое изложение этих вопросов, например, в книге [3].

Из таблицы 1 видно, что после 12-ступенной темперации лучшими являются 29-ступенная, 41-ступенная и 53-ступенная, так как в них натуральная квинта приближается лучше, чем в других. Важно отметить, что, как можно посчитать, лучшим следующим после числа 31/53 рациональным приближением числа $\alpha$ является число 210/359. Этот факт говорит нам о том, что за границами 53-ступенной темперации лучшие темперации нас не ожидают, если, кроме неудобства работы с большими числами ступеней, еще принять во внимание физиологию. А именно, тот физиологический факт, что человеческое ухо в среднем способно достаточно свободно различать музыкальные интервалы типа пифагоровой коммы, и сталкивается с трудностями в различии существенно меньших интервалов. (Отметим эксперименты по определению «зонной природы музыкального слуха», выполненных в 50-е годы Н.А. Гарбузовым ([4]), результаты которых практически совпадают с современными результатами по установлению частотных дифференциальных порогов слуха ([1]). В частности, были установлены физиологические пределы различимости: 5-20 c – для основных интервалов в частотах до 1 кГц в различных режимах восприятия).

Числа q = 12, 17, 29, 41 и 53 выступают здесь (в таблице 1) как возможные основания системы темперации. Мы видим, что чистая квинта в 53-ступенной темперации приближается лучше, чем в 41-ступенной в 7 раз, чем в 29-ступенной – в 22 раз, чем в 17-ступенной – в 58 раз, чем в 12-ступенной – в 29 раз. Таким образом, по отношению к чистой квинте 53-ступенная темперация в разы лучше других.

Рассмотрим теперь положение дел у 53-ступенной темперации с приближением обертонов. Напомним, что высоты тонов, задающих интервалы темперированного 12-звучного звукоряда, равны соответственно: 100 c (м. секунда), 200 c (б. секунда), 300 c (м. терция), 400 c (б. терция), 500 c (кварта), 600 c (тритон), 700 c (квинта), 800 (м. секста), 900 c (б. секста), 1000 c (м. септима), 1100 c (б. септима), 1200 c (октава). Обозначение c (англ. cent) означает центы или цент, интервал, равный 1/100 темперированного по 12-ступенной темперации полутона по логарифмической шкале, то есть интервал высоты $\log_2(\omega/\omega_1) = 1/1200$ (или частоты $\omega=\omega_1\, 2^{1/1200}$). Элементарный интервал в 53-ступенной темперированной системе = 1200/53 = 22,641509434 c (т.е. примерно равен пифагоровой комме 23,46 c).

*3. Высоты обертонов.* Пусть $\omega_1$– частота основного тона (в герцах = гц), тогда $\omega_k = k\,\omega_1$– частота k-го обертона (в гц). Поскольку высота чистого тона частоты $\omega$ по отношению к основному тону определяется формулой $h=\log_2(\omega/\omega_1)$ (см., напр., [1]), высоту (или интервальный коэффициент) $h_k$ k-го обертона вычисляем по формуле $h_k = \log_2(\omega_k/\omega_1) = \log_2 k = \ln k/ \ln 2$. Теперь, отбрасывая целую часть числа $h_k$ (число октав), выделим, таким образом, его мантиссу $\{h_k\}$ (дробную часть исходного логарифма) и представим её в центах, умножая на 1200. Результаты вычислений помещены в следующей таблице. Число $h_n = 1200(n-1)/53$ является высотой n-й ступени в центах.

*Таблица 2. Высоты обертонов по логарифмической шкале*

| Обертоны | k | Значение высоты $h_k$ | Мантисса $1200\{h_k\}$ высоты $h_k$ в центах | Ближайшая 53-ступень n | Отклонение $(\{h_k\}-h_n)\cdot\cdot1200$ в центах |
|---|---|---|---|---|---|
| Прима | 1 | 0,0000000000 | 0,0000000000 | 1 | 0 |
| Октава | 2 | 1,0000000000 | 0,0000000000 | 54 | 0 |
| Квинта | 3 | 1,5849625007 | 701,95500084 | 32 | 0,068208 |
| Большая терция | 5 | 2,3219280949 | 386,31371388 | 18 | 1,4081 |
| Малая терция | (5-3) | | 315,64128696 | 15 | –1,3398 |
| Малая септима | 7 | 2,8073549220 | 968,82590640 | 44 | –4,7590 |
| Большая секунда | 9 | 3,1699250014 | 203,91000168 | 10 | 0,13642 |
| (Ум. квинта) | 11 | 3,4594316186 | 551,31794232 | 25 | 7,9217 |
| (М. секста) | 13 | 3,7004397181 | 840,52766172 | 38 | 2,7918 |
| (Б. септима) | 15 | 3,9068905956 | 1088,2686672 | 49 | 1,4762 |
| (М. секунда) | 17 | 4,0874628416 | 104,95540992 | 6 | –8,2521 |
| (М. терция) | 19 | 4,2479275134 | 297,51301608 | 14 | 3,1734 |

Из этой таблицы видно, что кроме квинтового обертона в 53-ступенной темперации очень хорошо приближаются натуральные большая и малая терции, большая секунда, малая секста, большая септима и малая терция как 19-й обертон, также хорошо приближается малая септима. Хуже других обертонов приближаются в 53-ступенной темперации, оставшиеся среди первых двадцати 11-й и 17-й обертоны. Их отклонения примерно равны 8 центам. Таким образом, и в смысле приближения основных обертонов 53-ступенная темперация выглядит весьма подходящей.

И все же число звуков в октаве, равное 53, кажется пугающе большим. Как с ними управляться, как можно было бы расположить эти ступени удобно для исполнителя, да и для теоретика, если уж вообще об этом говорить? И не лучше ли ограничиться пусть худшими приближениями звуков квинтового (пифагорейского) и натурального звукорядов, но зато, казалось бы, более удобной темперацией с меньшим числом ступеней?

Ниже мы как раз и предлагаем такое размещение ступеней 53-ступенного темперированного звукоряда, которое представляется нам таким, которое не только основано на исторической преемственности имеющихся темперированных музыкальных систем (формула 5+7=12 реализована в фортепианной клавиатуре), но и естественно их развивает.

*4. Размещение ступеней звукоряда на трёх-мануальной клавиатуре.* Ниже предлагается несколько вариантов размещения ступеней 53-ступенного звукоряда на клавиатуре с тремя мануалами.

*Таблица 3. Размещение ступеней в 53-темперации*

1-й вариант

| | | | | | | | | | | | | | |
|---|---|---|---|---|---|---|---|---|---|---|---|---|---|
| 2 | 7 | 8 | 16 | 17 | 21 | 24 | 29 | 30 | 38 | 39 | 47 | 48 | 52 |
| | 3 | 11 | 12 | 20 | | | 25 | 33 | 34 | 42 | 43 | 51 | |
| 1 | 5 | 6 | 14 | 15 | 19 | 23 | 27 | 28 | 36 | 37 | 45 | 46 | 50 |
| | | 10 | | | | | | 32 | | 41 | | | |
| 53 | 4 | | 13 | | 18 | 22 | 26 | | 35 | | 44 | | 49 |
| | | 9 | | | | | | 31 | | 40 | | | |

2-й вариант

| | | | | | | | | | | | | | |
|---|---|---|---|---|---|---|---|---|---|---|---|---|---|
| 3 | 8 | 9 | 17 | 18 | 22 | 25 | 30 | 31 | 39 | 40 | 48 | 49 | 53 |
| | 4 | 12 | 13 | 21 | | | 26 | 34 | 35 | 43 | 44 | 52 | |
| 1 | 5 | 6 | 14 | 15 | 19 | 23 | 27 | 28 | 36 | 37 | 45 | 46 | 50 |
| | | 10 | | | | | | 32 | | 41 | | | |
| 2 | 7 | | 16 | | 20 | 24 | 29 | | 38 | | 47 | | 51 |
| | | 11 | | | | | | 33 | | 42 | | | |

3-й вариант

| | | | | | | | | | | | | | |
|---|---|---|---|---|---|---|---|---|---|---|---|---|---|
| 2 | 6 | 9 | 15 | 18 | 22 | 24 | 28 | 31 | 37 | 40 | 46 | 49 | 53 |
| | 4 | 11 | 13 | 20 | | | 26 | 33 | 35 | 42 | 44 | 51 | |
| 3 | 7 | 8 | 16 | 17 | 21 | 25 | 29 | 30 | 38 | 39 | 47 | 48 | 52 |
| | | 12 | | | | | | 34 | | 43 | | | |
| 1 | 5 | | 14 | | 19 | 23 | 27 | | 36 | | 45 | | 50 |
| | | 10 | | | | | | 32 | | 41 | | | |

Как вероятно уже догадался читатель, черные прямоугольники и квадраты означают здесь черные клавиши, а белые прямоугольники и многоугольники – белые клавиши типа фортепианных. Числа на них – номера соответствующих ступеней 53-ступенного темперированного звукоряда. Напомним, что элементарный (наименьший ненулевой) интервал в 53-ступенной системе = 22,641509433962 с и он примерно равен пифагоровой комме (23,46 с). Поэтому, если считать элементарный интервал равным пифагоровой комме (ниже для элементарного интервала в этой системе используется сокращение П, означающем большую греческую букву «Пи» в честь Пифагора), то интервалом между, скажем, *Cis* и *Des* будет как раз П. В интервал между, скажем, *C* и *Cis* попадает 5П, а между *C* и *Des* – 4П. То же справедливо и для других нот: *D* - *Dis* или *F-Fis* – это 5П, *D-Fes* – это 4П, а целый тон содержит 9П. Отметим также формулу, упомянутую еще в [10], стр.124 в связи с пифагорейским звукорядом: $5 \times 9 + 2 \times 4 = 53$. Поскольку 53-темперированную квинту (содержащую 31 П) с большой точностью можно считать чистой (натуральной) квинтой, то и саму 53-ступенную темперированную систему можно считать системой, отражающей звукоряд Пифагора, который, напомним, построен по чистым квинтам и октавам и продолжает реальную диатонику в область хроматики.

Теперь посмотрим на расположение диатонических интервалов и обертонов от основного тона в каком-то из представленных выше вариантов размещения ступеней. Вспомним, что в пятом столбце таблицы 2 указаны ступени 53-ступенного темперированного звукоряда, близкие к обертонам от основного тона, и изобразим их на клавиатуре. Расположение диатонических интервалов и ступеней натурального ряда от ноты *C* на клавиатуре, для примера, из 3-го варианта расположения дано в таблице 4, что фактически реализует результаты вычислений из пятого столбца таблицы 2. Ниже, в п.6, обсуждается расположение ступеней натурального ряда от ноты *C* в 1-м варианте размещения, которое представляется более удачным. Ниже под диатоническим интервалом, как это общепринято, понимается интервал, построенный в пифагорейской (т.е. квинтовой) системе.

*Таблица 4. Расположение диатонических интервалов и ступеней натурального ряда*

(3-й вариант)

| | | | | | | | | | | | | | |
|---|---|---|---|---|---|---|---|---|---|---|---|---|---|
| | О.М. секу-нда 6 | 9 | О.М. тер-ция 15 | О.Б. тер-ция 18 | | | Д.Ув. квар-та 28 | 31 | 37 | 40 | 46 | О.Б. сеп-тима 49 | |
| 2 | 4 | 11 | 13 | 20 | 22 | 24 | 26 | 33 | 35 | 42 | О.М. сеп-тима 44 | 51 | 53 |
| | 7 | 8 | 16 | 17 | | | 29 | О.Ув квар-та 30 | О.М.с екс-та 38 | 39 | 47 | 48 | |
| 3 | | 12 | | 21 | | О. Ум. квинта 25 | | 34 | | 43 | | 52 | |
| | Д. М. секунда 5 | | О.М. тер-ция (19-й обертон) 14 | | | | Д.Ум. квинта 27 | | Д. М. секста 36 | | Д. М. септима 45 | | |
| **Основной тон** Прима 1 | | Д. и О. Б. секунда 10 | | Д. Б. терция 19 | | Д. и О. кварта 23 | | Д. и О. квинта 32 | | Д. Б. секста 41 | | Д. Б. септима 50 | |

В таблице 4 используются сокращения: Д. – диатоническая или диатонический, О. – обертоновая (натуральная), Б. – большая, М. – малая, Ув. — увеличенная, Ум. – уменьшенная. Разумеется, правомерность использования нами термина «обертоновый интервал» (за неимением лучшего, установившегося) может вызвать дискуссию.

*5. Некоторые выводы.* Рассмотрим таблицу 3. Из нее видно, что во всех трех вариантах расположения ступеней нижний мануал настроен почти так же, как и в 12-ступенной системе. Правда, с теми различиями, что в первом и третьем вариантах расположения: 1) целый тон, например, (1-10), здесь равен 9П=203,77358591 с, в 12-ступенной – 200; 2) полутоны восходящие: с белой вверх – 90,56603773 с, а с черной вверх – 113,20754714 с, а во втором варианте – наоборот.

Средний мануал во всех трех вариантах расположения ступеней отвечает 17-звучной системе интонирования в тональности до-мажор (ля-минор), где имеются клавиши для звуков с пятью диезами и пятью бемолями, но для дубль-диеза или дубль-бемоля нужно выходить за рамки среднего мануала. Верхний мануал в третьем варианте расположения ступеней отвечает 24-звучной четвертитоновой системе Алоиса Хаба, а в двух других является набором хроматизмов.

Запишем еще следующие важные равенства:

$$5+7=12,\ 12+5=17,\ 12+12=24,\ 12+17=29,\ 12+29=41,\ 12+17+24=53, \qquad (*)$$

которые отмечались многими авторами. Последние три из них для нас означают, что при наличии известных технических средств, мы можем темперировать пару стоящих рядом мануалов как 29- или 41- ступенную темперацию, оставляя неиспользованный мануал без внимания или используя его для других тембров. Отметим также, что темперированную систему А. Хаба мы можем получить при наличии технических средств, забыв о двух нижних мануалах и темперируя один верхний. Точно так же мы можем темперировать средний мануал по 17-звучной системе или (и) нижний по 12-звучной.

Приведем для полноты изложения также и варианты размещения ступеней на двух мануалах предлагаемой нами трех-мануальной системы для случаев 29-ступенной (два нижних мануала) и 41-ступенной темпераций (два верхних мануала).

*Таблица 5. Размещение ступеней в 29-темперации на двух мануалах*

1-й вариант

| | | | | | | | | | | | | | |
|---|---|---|---|---|---|---|---|---|---|---|---|---|---|
| 1 | 3 | 4 | 8 | 9 | 11 | 13 | 15 | 16 | 20 | 21 | 25 | 26 | 28 |
| | | 6 | | | | | | 18 | | 23 | | | |
| 2 | 5 | | 10 | | 12 | 14 | 17 | | 22 | | 27 | | 29 |
| | | 7 | | | | | | 19 | | 24 | | | |

2-й вариант

| | | | | | | | | | | | | | |
|---|---|---|---|---|---|---|---|---|---|---|---|---|---|
| 2 | 4 | 5 | 9 | 10 | 12 | 14 | 16 | 17 | 21 | 22 | 26 | 27 | 29 |
| | | 7 | | | | | | 19 | | 24 | | | |
| 1 | 3 | | 8 | | 11 | 13 | 15 | | 20 | | 25 | | 28 |
| | | 6 | | | | | | 18 | | 23 | | | |

Здесь следует отметить, что в 29-ступенной темперации минимальный интервал равен 41,38 с, что сразу указывает на существенную трудность, связанную с сильно завышенной (по отношению к пифагоровой комме) разницей между рядом стоящими диезной и бемольной ступенями (например, Db-C#).

*Таблица 6. Размещение ступеней в 41-темперации на двух мануалах*

1-й вариант

| | | | | | | | | | | | | | |
|---|---|---|---|---|---|---|---|---|---|---|---|---|---|
| 2 | 6 | 7 | 13 | 14 | 17 | 19 | 23 | 24 | 30 | 31 | 37 | 38 | 41 |
| | 3 | 9 | 10 | 16 | | | 20 | 26 | 27 | 33 | 34 | 40 | |
| 1 | 4 | 5 | 11 | 12 | 15 | 18 | 21 | 22 | 28 | 29 | 35 | 36 | 39 |
| | | 8 | | | | | | 25 | | 32 | | | |

2-й вариант

| | | | | | | | | | | | | | |
|---|---|---|---|---|---|---|---|---|---|---|---|---|---|
| 1 | 4 | 5 | 11 | 12 | 16 | 18 | 21 | 22 | 28 | 29 | 35 | 36 | 40 |
| | 2 | 8 | 9 | 15 | | | 19 | 25 | 26 | 32 | 33 | 39 | |
| 3 | 6 | 7 | 13 | 14 | 17 | 20 | 23 | 24 | 30 | 31 | 37 | 38 | 41 |
| | | 10 | | | | | | 27 | | 34 | | | |

Здесь также следует отметить, что в 41-ступенной темперированной системе минимальный интервал равен 29,27 с, что приближает пифагорову комму (23,46 с, например, *Des-Cis*) лучше, нежели в 29-системе (41,38 с), но все же хуже, чем в 53-системе (22,64 с).

*6. Кварто-квинтовый круг и системы интонирования.* Рассмотрим теперь кварто-квинтовый круг для 53-звучной темперированной системы, вводя обозначения: *###* – три диеза, *bbb* – три бемоля, *4#* – четыре диеза и *4b* –четыре бемоля.

*Таблица 7. Кварто-квинтовый круг*

1-й вариант

| | | | | | | | | | | | | | |
|---|---|---|---|---|---|---|---|---|---|---|---|---|---|
| 2 | 7<br>H## | 8<br>Fbbb | 16<br>C### | 17<br>Gbbb | 21 | 24 | 29<br>E## | 30<br>D4#=H4b | 38<br>F### | 39<br>Cbbb | 47<br>G### | 48<br>Dbbb | 52 |
| H# | 3<br>A### | 11<br>C## | 12<br>H### | 20<br>D## | C4#=<br>A4b | E# | 25<br>D### | 33<br>F## | 34<br>E### | 42<br>G## | 43<br>F4#=D4b | 51<br>A## | G4#=<br>E4b |
| 1 | 5<br>Db | 6<br>C# | 14<br>Eb | 15<br>D# | 19 | 23 | 27<br>Gb | 28<br>F# | 36<br>Ab | 37<br>G# | 45 B<br>=Hb | 46<br>A# | 50 |
| C | | 10<br>D | | E | | F | | 32<br>G | | 41<br>A | | H | |
| 53 | 4<br>Ebbb | | 13<br>Fbb | | 18 | 22 | 26<br>Abbb | | 35<br>Hbbb | | 44<br>Cbb | | 49 |
| Dbb | | 9<br>Ebb | | Fb | | Gbb | | 31<br>Abb | | 40<br>Hbb | | Cb | |

2-й вариант

| | | | | | | | | | | | | | |
|---|---|---|---|---|---|---|---|---|---|---|---|---|---|
| 3 | 8<br>Fbbb | 9<br>Ebb | 17<br>Gbbb | 18<br>Fb | 22 | 25 | 30D4#=<br>H4b | 31<br>Abb | 39<br>Cbbb | 40<br>Hbb | 48<br>Dbbb | 49<br>Cb | 53 |
| A### | 4<br>Ebbb | 12<br>H### | 13<br>Fbb | 21C4#=<br>A4b | Gbb | D### | 26<br>Abbb | 34<br>E### | 35<br>Hbbb | 43F4#=<br>D4b | 44<br>Cbb | 52G4#=<br>E4b | Dbb |
| 1 | 5<br>Db | 6<br>C# | 14<br>Eb | 15<br>D# | 19 | 23 | 27<br>Gb | 28<br>F# | 36<br>Ab | 37<br>G# | 45 B<br>=Hb | 46<br>A# | 50 |
| C | | 10<br>D | | E | | F | | 32<br>G | | 41<br>A | | H | |
| 2 | 7<br>H## | | 16<br>C### | | 20 | 24 | 29<br>E## | | 38<br>F### | | 47<br>G### | | 51 |
| H# | | 11<br>C## | | D## | | E# | | 33<br>F## | | 42<br>G## | | A## | |

3-й вариант со смещением номеров ступеней на 2 вниз

| | | | | | | | | | | | | | |
|---|---|---|---|---|---|---|---|---|---|---|---|---|---|
| 53 | 4<br>Ebbb | 7<br>H## | 13<br>Fbb | 16<br>C### | 20 | 22 | 26<br>Abbb | 29<br>E## | 35<br>Hbbb | 38<br>F### | 44<br>Cbb | 47<br>G### | 51 |
| Dbb | 2<br>H# | 9<br>Ebb | 11<br>C## | 18<br>Fb | D## | Gbb | 24<br>E# | 31<br>Abb | 33<br>F## | 40<br>Hbb | 42<br>G## | 49<br>Cb | A## |
| 1 | 5<br>Db | 6<br>C# | 14<br>Eb | 15<br>D# | 19 | 23 | 27<br>Gb | 28<br>F# | 36<br>Ab | 37<br>G# | 45 B<br>=Hb | 46<br>A# | 50 |
| C | | 10<br>D | | E | | F | | 32<br>G | | 41<br>A | | H | |
| 52 | 3<br>A### | | 12<br>H### | | 17 | 21 | 25<br>D### | | 34<br>E### | | 43<br>F4#=D4b | | 48 |
| G4#=E4b | | 8<br>Fbbb | | Gbbb | | C4#=A4b | | 30<br>D4#=H4b | | 39<br>Cbbb | | Dbbb | |

Следует отметить, что во всех трех вариантах расположения ступеней имеются одни и те же четыре энгармонические приравнивания: *G4#=E4b, C4#=A4b, D4#=H4b, F4#=D4b,* в которых только *D* встречается дважды, а остальные – по разу. Конечно, это не случайно и может быть объяснено. Отметим также, что других четырехкратных диезов или бемолей больше нет.

Наблюдая расположение ступеней в приведенных вариантах, можно сделать выводы о возможных упрощениях в использовании клавиатуры для исполнения тех или иных музыкальных произведений. Так, например, если в тексте исполняемого произведения, богатого на хроматизмы, отсутствуют двойные и выше диезы, а также *H#* и *E#,* но присутствуют бемольные понижения, то можно воспользоваться 1-м вариантом клавиатуры с использованием двух нижних мануалов:

*Таблица 8. 29-звучная пифагорейская система как часть 53-ступенной темперации*

Вариант 1. Сдвиг в бемольную сторону.

| | | | | | | | | | | | | | |
|---|---|---|---|---|---|---|---|---|---|---|---|---|---|
| 1 | 5 Db | 6 C# | 14 Eb | 15 D# | 19 | 23 | 27 Gb | 28 F# | 36 Ab | 37 G# | 45 B =Hb | 46 A# | 50 |
| C | 10 D | E | F | 32 G | 41 A | H | | | | | | | |
| 53 | 4 Ebbb | 13 Fbb | 18 | 22 | 26 Abbb | 35 Hbbb | 44 Cbb | 49 | | | | | |
| Dbb | 9 Ebb | Fb | Gbb | 31 Abb | 40 Hbb | Cb | | | | | | | |

Вариант 2. Сдвиг в диезную сторону.

| | | | | | | | | | | | | | |
|---|---|---|---|---|---|---|---|---|---|---|---|---|---|
| 1 | 5 Db | 6 C# | 14 Eb | 15 D# | 19 | 23 | 27 Gb | 28 F# | 36 Ab | 37 G# | 45 B =Hb | 46 A# | 50 |
| C | 10 D | E | F | 32 G | 41 A | H | | | | | | | |
| 2 | 7 H## | 16 C### | 20 | 24 | 29 E## | 38 F### | 47 G### | 51 | | | | | |
| H# | 11 C## | D## | E# | 33 F## | 42 G## | A## | | | | | | | |

Отметим также, что в этом случае на клавиатуре варианта 1 присутствуют клавиши, ответственные за ступени обертонового звукоряда от ноты *C*: 18-я (5-й обертон – б. терция), 15-я (интервал между 5-м и 3-м обертонами – м.терция), 44-я (7-й обертон – м. септима), 10-я (9-й обертон – б. секунда), 49-я (15-й обертон – б. септима), 6-я (17-й обертон, м. секунда) и 14-я (19-й обертон, м. терция), но отсутствуют 25-я (11-й обертон – ум. квинта) и 38-я (13-й обертон – м. секста), отсутствующие клавиши находятся на верхнем мануале. Кстати сказать, тоника параллельной тональности к C-dur приходится, похоже, скорее не на 41, а 40-ю клавишу, т.к. тогда именно на 1-ю клавишу приходится обертон малой терции 5-3 от звука 40-й клавиши.

Если же в тексте исполняемого произведения, богатого на хроматизмы, отсутствуют двойные и выше бемоли, а также *Cb* и *Fb*, но присутствуют диезные повышения, то можно воспользоваться 2-м вариантом таблицы 8 (и таблицы 7) клавиатуры с использованием двух нижних мануалов.

Наконец, два верхних мануала 3-го варианта расположения таблицы 7 можно использовать для исполнения возможных произведений в 41-звучной системе, в которой отсутствуют редкие ступени *A###, H###, D###, E###, Fbbb, Gbbb, Cbbb, Dbbb, G4#=E4b, C4#=A4b, D4#=H4b, F4#=D4b* на двух верхних мануалах**:**

*Таблица 9. 41-звучная пифагорейская система как часть 53-ступенной темперации*

| | | | | | | | | | | | | | |
|---|---|---|---|---|---|---|---|---|---|---|---|---|---|
| 53 | 4<br>Ebbb | 7<br>H## | 13<br>Fbb | 16<br>C### | 20 | 22 | 26<br>Abbb | 29<br>E## | 35<br>Hbbb | 38<br>F### | 44<br>Cbb | 47<br>G### | 51 |
| Dbb | 2<br>H# | 9<br>Ebb | 11<br>C## | 18<br>Fb | D## | Gbb | 24<br>E# | 31<br>Abb | 33<br>F## | 40<br>Hbb | 42<br>G## | 49<br>Cb | A## |
| 1 | 5<br>Db | 6<br>C# | 14<br>Eb | 15<br>D# | 19 | 23 | 27<br>Gb | 28<br>F# | 36<br>Ab | 37<br>G# | 45 в=<br>Hb | 46<br>A# | 50 |
| C | | 10<br>D | | E | | F | | 32<br>G | | 41<br>A | | H | |

Интересно отметить, что здесь присутствуют клавиши, ответственные за все (до 19-й) ступени обертонового звукоряда от ноты *C* таблицы 2, кроме 25-й (ум. квинты).

В вышеприведенных расстановках ступеней основным считался средний мануал, в который наряду с семью диатоническими ступенями помещены альтерации пяти из них. Это может показаться неудобным, поэтому ниже предлагается три других варианта размещения ступеней, где основным является нижний мануал, а два других рассматриваются как вместилище хроматизмов. Заинтересованный читатель может сам проанализировать эти размещения и составить о них свое мнение.

*Таблица 10. Размещение ступеней в 53-темперации (продолжение)*

4-й вариант

| | | | | | | | | | | | | | |
|---|---|---|---|---|---|---|---|---|---|---|---|---|---|
| 2 | 7 | 8 | 16 | 17 | 21 | 24 | 29 | 30 | 38 | 39 | 47 | 48 | 52 |
| | 3 | 11 | 12 | 20 | | | 25 | 33 | 34 | 42 | 43 | 51 | |
| | 4 | 6 | 13 | 15 | | | 26 | 28 | 35 | 37 | 44 | 46 | |
| 53 | | 9 | | 18 | | 22 | | 31 | | 40 | | 49 | |
| 1 | 5 | | 14 | | 19 | 23 | 27 | | 36 | | 45 | | 50 |
| | | 10 | | | | | | 32 | | 41 | | | |

5-й вариант

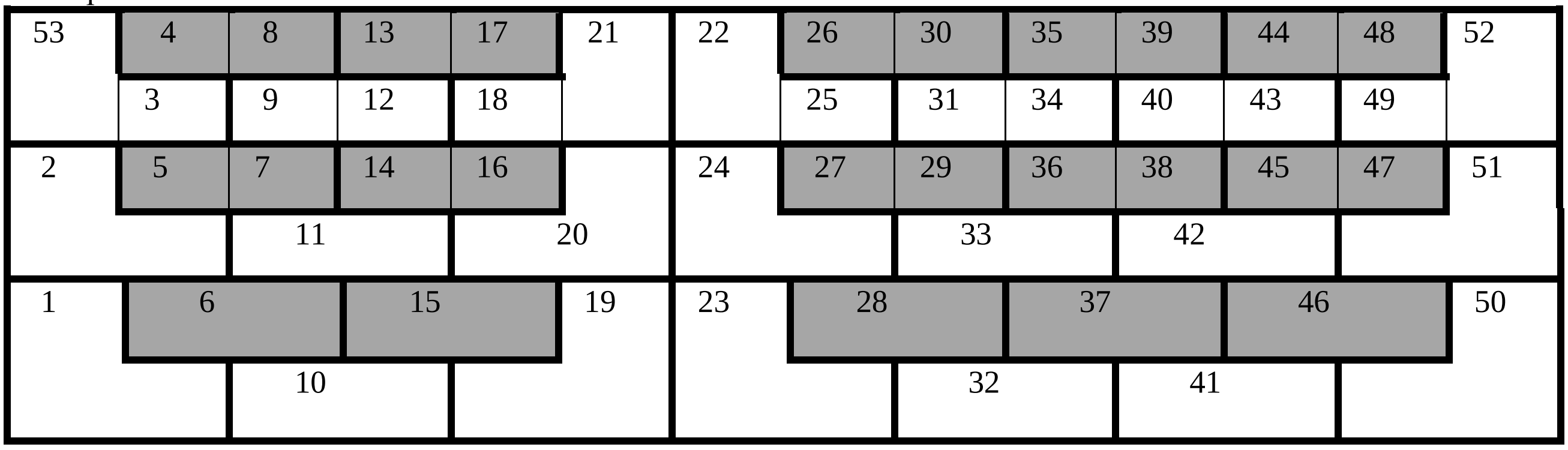

6-й вариант

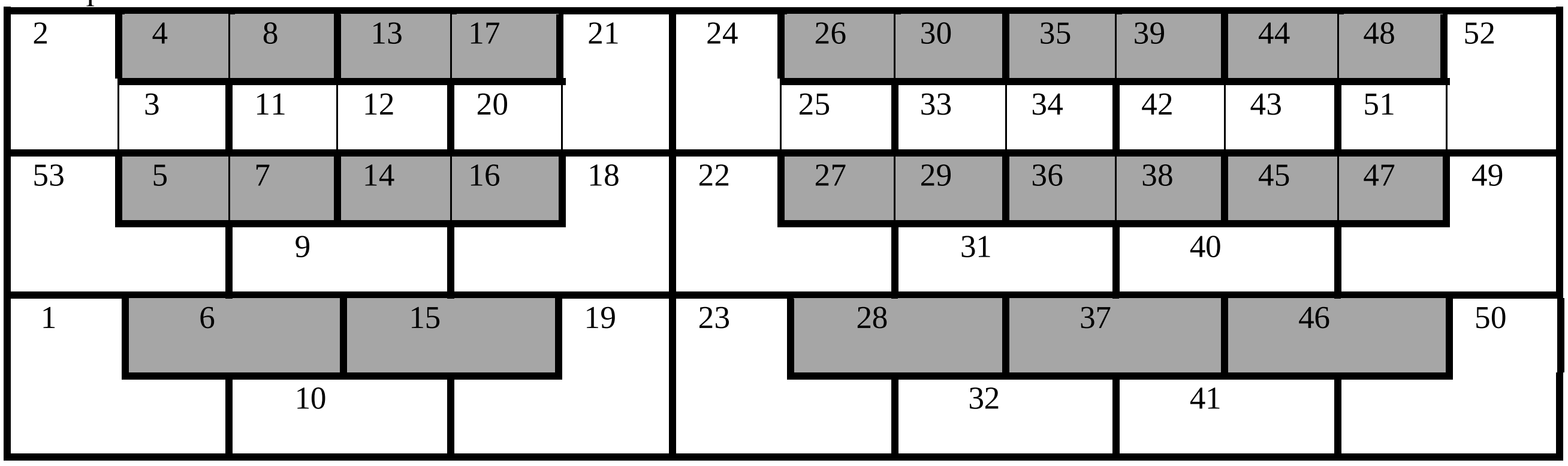

*7. Заключительные замечания.* Выше были представлены аргументы в пользу 53-ступенной темперации и предложены способы расстановки ступеней этой системы на трех-мануальной клавиатуре. В частности, реализована формула, отмеченная в [10]: $5 \times 9 + 2 \times 4 = 53$, где 9П = тон, а 4П = полутон *E-F* и *H-C*.

Следует сказать, что способы расстановки ступеней не исчерпываются приведенными выше в таблицах 3 и 10, и читатель, при желании, может сам поупражняться в расстановках ступеней, добиваясь нужного ему результата. Подобные упражнения могут быть полезными также в 29-ти и 41-звучных темперированных системах. Понятно, что при развитии в этом направлении привлечения и использования иных, кроме 12-звучной, темперированных систем музыкальная теория начнет приобретать некоторые совершенно иные, незнакомые нам пока черты, хотя, конечно, базовые положения теории измениться не могут.

При рассмотрении возможностей развития музыкальной теории и практики посредством привлечения других темперированных систем, возникает огромное количество различных вопросов теоретического и практического характера. Так, перед разработчиками и изготовителями инструмента, реализующего предложенную клавиатуру, особенно в части верхнего мануала, встает проблема мензуры (геометрических параметров) клавиатуры.

В теоретическом плане, например, встают многочисленные вопросы о нотации. Как рационально изображать обычный для исполнителя музыкальный текст так, чтобы и привлечь какие-то новые открывающиеся возможности исполнения и интонирования хроматизмов, и не утонуть в этом многообразии ступеней? Или: как записать народную музыку, основанную на иных звуковых шкалах и системах, например, использующих 17 ступеней (азербайджанская музыка, интервалы в 90с$\approx$4П, 204с$\approx$9П и т.д.) или 22 (индийская музыка, наименьший интервал $\approx$ 3П) ступени неравномерных темпераций? Разработка этих вопросов ведется в различных направлениях (см., например, ссылки на работы по турецкой монодии в статье [14]), но при использовании предлагаемой выше клавиатуры появляется еще одна возможность для графической записи звуков. Это – возможность использовать общепринятую нотацию для изображения звуков на письме, просто привлекая цвета для обозначения нот на мануалах и раскрашивая ноты на одном и том же мануале в один и тот же цвет. Напрашивается, например, менять положение ноты относительно штиля (слева или справа от штиля) или оттенок цвета ноты на нотоносце для обозначения левой или правой клавиши. При этом записанная нота будет обозначать уже не звук определенной высоты, а клавишу, и тогда всем надо заранее договариваться, в какой из систем расстановки ступеней записано то или иное музыкальное произведение.

Автор считает своим долгом отметить, что настоящая работа содержит реализацию идей и помыслов, витавших во время многочисленных и продолжительных обсуждений и бесед о музыке, её настоящем и будущем, в частности, при обсуждении книги [6] и вышеприведенных равенств (*). Эти обсуждения происходили в начале 1980-х годов в Донецке во время встреч, в которых участвовали, как правило, трое человек: преподаватель музыкально-теоретических дисциплин Николай Константинович Шевченко, член Союза композиторов Украины Евгений Григорьевич Милка и я, в прошлом музыкант, а ныне математик и автор настоящей статьи, в которой реализованы некоторые из тех витавших в воздухе идей. К сожалению, эта реализация затянулась. Саму статью автор посвящает светлой памяти выдающихся музыкантов Н.К. Шевченко и Е.Г. Милки.

Автор выражает свою признательность проф. Некрасову А.И. и проф. Ивко В.Н. за обсуждения и особую благодарность доктору Лыжову Г.И. за важные замечания и помощь.

В заключение автор хотел бы выразить надежду, что читатель нашел или еще найдет в будущем интересными идеи или схемы, изложенные в настоящей статье.

*Литература*